# First-principles study of anisotropic thermoelectric transport properties of IV-VI semiconductor compounds SnSe and SnS

Ruiqiang Guo[1], Xinjiang Wang[1], Youdi Kuang[1] and Baoling Huang[1,2*]

[1]*Department of Mechanical and Aerospace Engineering, The Hong Kong University of Science and Technology, Clear Water Bay, Kowloon, Hong Kong*

[2]*The Hong Kong University of Science and Technology Shenzhen Research Institute, Shenzhen, 518057, China*

Abstract

Tin selenide (SnSe) and tin sulfide (SnS) have recently attracted particular interest due to their great potential for large-scale thermoelectric applications. A complete prediction of the thermoelectric performance and the understanding of underlying heat and charge transport details are the key to further improvement of their thermoelectric efficiency. We conduct comprehensive investigations of both thermal and electrical transport properties of SnSe and SnS using first-principles calculations combined with the Boltzmann transport theory. Due to the distinct layered lattice structure, SnSe and SnS exhibit similarly anisotropic thermal and electrical behaviors. The cross-plane lattice thermal conductivity $\kappa_L$ is 40-60% lower than the in-plane values. Extremely low $\kappa_L$ is found for both materials because of high anharmonicity while the average $\kappa_L$ of SnS is ~ 8% higher than that of SnSe from 300 to 750 K. It is suggested that nanostructuring would be difficult to further decrease $\kappa_L$ because of the short mean free paths of dominant phonon modes (1-30 nm at 300 K) while alloying would be efficient in reducing $\kappa_L$ considering that the relative $\kappa_L$ contribution (~ 65%) of optical phonons is remarkably large. On the electrical side, the anisotropic electrical conductivities are mainly due to the different effective masses of holes and electrons along the *a*, *b* and *c* axes. This leads to the highest optimal *ZT* values along the *b* axis and lowest ones along the *a* axis in both *p*-type materials. However, the *n*-type ones exhibit the highest *ZT*s along the *a* axis due to the enhancement of power factor when the chemical potential gradually approaches the secondary conduction band valley that causes significant increase in electron mobility and density of states. Owing to the larger mobility and smaller $\kappa_L$ along the given direction, SnSe exhibits larger optimal *ZT*s compared with SnS in both *p*- and *n*-type materials. For both materials, the peak *ZT*s of *n*-type materials are much higher than those of *p*-type ones along the same direction. The predicted highest *ZT* values at 750 K are 1.0 in SnSe and 0.6 in SnS along the *b* axis for the *p*-type doping while those for the *n*-type doping reach 2.7 in SnSe and 1.5 in SnS along the *a* axis, rendering them among the best bulk thermoelectric materials for large-scale applications. Our calculations show reasonable agreements with the experimental results and quantitatively predict the great potential in further enhancing the thermoelectric performance of SnSe and SnS, especially for the *n*-type materials.

# I. INTRODUCTION

Thermoelectric materials, which enable direct interconversion between thermal and electrical energy, are being actively considered for a variety of energy harvesting and thermal management applications[1-4]. The efficiency of thermoelectric materials is often measured by a dimensionless figure of merit $ZT = S^2\sigma T/(\kappa_L+\kappa_e)$, where $S$ is the Seebeck coefficient, $\sigma$ is the electrical conductivity, $T$ is the temperature, and $\kappa_L$ and $\kappa_e$ are the phononic and electronic contributions to the thermal conductivity, respectively. On searching for high performance thermoelectric materials, high $ZT$s have been continuously reported for the IV-VI compounds such as PbTe[4-7], PbSe[8-10], GeTe[11-13], SnTe[14-17] and their alloys[1,18]. These semiconductors are of high symmetric rock salt structure ($Fm\bar{3}m$ group) and possess isotropic properties. Another set of IV-VI compounds, such as SnSe and SnS, which crystalize in highly anisotropic layered orthorhombic structure (*Pnma* group) at room temperature and transit into the *Cmcm* group around 800 K[19-21], are attracting great interest for their potential in thermoelectric applications. Figure 1 shows the lattice structure (*Pnma* group) of SnSe (SnS) with an orthorhombic unit cell consisting of eight atoms, which form double layers (*b-c* plane) that are perpendicular to the *a*-axis. Sn and Se (S) atoms within the layers are covalently bonded and construct zig-zagging chains along the *b*-axis while those between the layers are weakly coupled through a van der Waals-like interaction.

Consisting of earth-abundant and low-cost elements, SnSe and SnS are very attractive for large-scale applications. Although their optical properties have been intensively investigated[22,23], their thermoelectric properties have long been ignored since a simple study in 1963, which reported room temperature $ZT$s of 0.15 for SnSe and 0.27 for SnS, respectively[24]. Recently, $ZT$s as high as 2.6 at 923 K for single crystalline *p*-type SnSe have stimulated renewed interests in SnSe and SnS for thermoelectric applications[19]. Such a high $ZT$ value is mainly due to a combination of its ultralow intrinsic lattice thermal conductivity and moderate power factor. Subsequent experimental investigations on *p*-type polycrystalline SnSe[20,25], SnS[21,26] and SnS$_{1-x}$Se$_x$[26], inexplicably, reported much lower *ZT*s at various temperatures. The large discrepancy between the single- and poly-crystalline counterparts promotes further researches on the thermoelectric properties of SnSe and SnS. Meanwhile, most previous experimental studies focus on *p*-type crystals and the thermoelectric properties of *n*-type SnSe or SnS are rarely reported due to the difficulty in synthesizing high-quality *n*-type crystals[27,28]. Due to the large scatter in experimental results and the difficulties in synthesizing and characterizing high-quality samples, it is desirable to accurately predict the individual thermoelectric parameters and develop a fundamental understanding of heat and charge transports in these highly anisotropic IV-VI compounds, which is important for the precise evaluation of the potential of SnSe and SnS-based materials in thermoelectric applications and may shed light on the further improvement of their performance.

A few first-principles papers have been conducted to investigate some separating properties of SnSe and SnS. Shi and Kioupakis and Kutorasinski *et al.* discussed the dependence of electrical properties on doping concentration and temperature in SnSe under the constant relaxation time approximation[29,30]. Carrete *et al.* applied first-principles to calculate the lattice thermal conductivity of pristine SnSe and reported significantly larger values than the experimental results of single crystals[31]. Parker and Singh calculated the band structure of SnS and predicted a high Seebeck coefficient[32]. Bera *et al.* reported that the power factor of SnS can be potentially improved by silver doping and the carrier concentration in SnSe can be further optimized despite its significant intrinsic *p*-type behavior[33]. However, due to the strong coupling between *S*, $\sigma$ and $\kappa$, e.g., doping may change these three parameters simultaneously, a complete prediction and optimization of the *ZT* value requires a comprehensive study of both thermal and electrical properties under the same conditions. To date, the theoretical optimal *ZT*s of these two semiconductors are still unknown. A detailed comparison of the thermoelectric transport properties between SnSe and SnS is still lacking. Meanwhile, the phonon transport details in SnS have rarely been theoretically explored. It is of significance to conduct a systematical investigation on both *p*- and *n*-type thermoelectric SnSe and SnS.

In this work, we investigate the thermoelectric transport properties of *p*- and *n*-type SnSe and SnS of the *Pnma* group with first-principles calculations. We first calculate the lattice thermal conductivities of both materials using first-principles and the Boltzmann transport equation (BTE). This approach[34-38] has recently been implemented on many semiconductors such as Si[37,39], $CoSb_3$[40], $Mg_2Si$[41], PbTe[42] and PbSe[42], showing good agreements with the experimental results. Detailed phonon spectral distributions and mode contributions to $\kappa_L$ are analyzed to understand the low anisotropic lattice thermal conductivity and the difference between SnSe and SnS. The mean free path (MFP) distributions are also analyzed for guiding the further reduction of $\kappa_L$ by multi-scale structural engineering. We then determine the electrical properties including the Seebeck coefficient, electrical conductivity, and electronic thermal conductivity using the BTE for charges[43,44]. Therefore, *ZT*s of *p*- and *n*-type SnSe and SnS at varying carrier concentrations and temperatures are obtained and compared. It is found that SnSe has larger optimal *ZT*s compared with SnS for both *p*- and *n*-type and both *n*-type materials possess much higher potential *ZT* values than the corresponding *p*-type ones, which are mainly due to the secondary band valley close to the band edge.

## II. PREDICTION OF PHONON TRANSPORT

### A. Calculation method

Based on the phonon BTE, the lattice thermal conductivity can be calculated by summing the contributions of all phonon modes denoted by the wavevector $\mathbf{q}$ and the dispersion branch $s$

$$\kappa_{L,\alpha\beta} = \sum_{\mathbf{q}s} C_V(\mathbf{q}s) v_g^\alpha(\mathbf{q}s) v_g^\beta(\mathbf{q}s) \tau_{\mathbf{q}s} , \qquad (1)$$

where $\kappa_L$ is a second-order tensor with the subscripts $\alpha$ and $\beta$ denoting its components, $C_V$ is the phonon mode volumetric specific heat, $v_g$ is the group velocity and $\tau$ is the phonon lifetime. To extract phonon dispersions and scattering rates, accurate harmonic and anharmonic interatomic force constants (IFCs) are required. In this study, the harmonic and third-order anharmonic IFCs were obtained from density functional theory (DFT) calculations, implemented with the projector augmented wave (PAW)[45] pseudopotentials using the Vienna Ab initio simulation package[46] (VASP). The local density approximation (LDA) of Perdew and Zunger[47], which is typically used for layered structures where van der Waals interactions are important[23], was used for the exchange-correlation functional of DFT. The conventional unit cells of SnSe and SnS were first fully relaxed with a 4×12×12 **k**-point sampling grid and a cut-off energy of 400 eV, which were tested to ensure the convergence. A 2×3×3 supercell was then constructed and used for the second-order and third-order IFCs calculations using the direct method[38], which involves the displacements of one and two atoms, respectively. The static first-principles calculations were conducted with a precision as high as $10^{-8}$ eV for the total energy difference between two successive self-consistency steps and $10^{-6}$ eV/Å as the convergence criterion for the forces on atoms. The phonon frequencies were then calculated by Fourier transformation based on the extracted harmonic IFCs while three-phonon scattering rates were determined by Fermi's golden rule with the anharmonic IFCs as input. A 11×27×27 **q** mesh for phonons was carefully checked to guarantee convergent results from 300 to 750 K. The three-phonon scattering rates, together with isotopic impurity scattering rates, were then used to solve the linearized phonon BTE with an iterative approach. The van der Waals corrections were found to overestimate the phonon frequencies of both materials and therefore were not applied in this work. More calculation details of this method can be found in our previous papers[39,40].

**B. Harmonic properties**

The relaxed lattice constants resulting from LDA are $a$ = 11.31 Å, $b$ = 4.12 Å and $c$ = 4.29 Å for SnSe, and $a$ = 10.97 Å, $b$ = 3.96 Å and $c$ = 4.16 Å for SnS, which are close to previous *ab initio* calculations[23,48,49] but are slightly underestimated compared with the experimental results[19,21]. This agrees with the overbinding effect of LDA. The LDA bulk modulus for SnSe and SnS are 39.4 and 41.6 GPa, respectively, in good agreement with the experimental data (35.0 GPa for SnSe[50] and 36.6 GPa for SnS[23]). The computed phonon dispersions of SnSe and SnS are shown in Figs. 2 (a) and (b). The longitudinal optical - transverse optical (LO-TO) splitting is predicted by incorporating the effects of long-range Coulomb interactions based on the Born effective charges and dielectric constants calculated by first-principles. SnSe and SnS show very similar dispersion curves along different high symmetry lines (refer to the Brillouin zone in Fig. 1) except that the frequencies of optical branches of SnS are significantly higher, which agrees with the experimental results[51] and is mainly due to the smaller atomic mass of S. The phonon dispersions are characterized by markedly dispersive optical branches, indicating significant group velocities that can be as large as those of acoustic phonons along the Γ→Y and Γ→Z directions. However, the group velocities of optical phonons along the Γ→X direction are much smaller, which can be attributed to the weak interactions between the layers. Significant group velocities of optical phonons were also reported in the rocksalt structure IV-VI group semiconductors such as PbSe and PbTe[42]. The sound velocities of SnSe are slightly smaller than those for SnS along the three lattice directions. Figures 2 (c) and (d) compare the phonon density of states of SnSe and SnS. One can find that the DOS are separated into two major regimes, with the lower and higher frequency parts mainly contributed by the vibrations of Sn and Se (S), respectively. Owing to the larger cation/anion mass ratio, there is almost a complete band gap from 4.1 to 4.4 THz for SnS, which is absent for SnSe.

**C. Lattice Thermal conductivity**

Figure 3 shows the temperature-dependent $\kappa_L$ of SnSe and SnS along the *a*, *b* and *c* axes, in comparison with the experimental data[19-21,24,25]. The lattice thermal conductivities of both materials are found to be very low and anisotropic. For example, the calculated room-temperature $\kappa_L$ of SnSe along the *a*, *b* and *c* axes are 0.8, 2.0 and 1.7 W/mK, respectively, even lower than that (2.2 W/mK) of PbSe consisting of heavier elements. As the temperature rises to 750 K, $\kappa_L$ reduces to 0.3, 0.8 and 0.7 W/mK along the *a*, *b* and *c* axes, respectively. Specifically, the value along the *a* axis approaches the predicted amorphous limit (0.26 W/mK)[52], which is extremely low for crystalline solids. Natural isotopes only lead to a small reduction in $\kappa_L$ for both materials, e.g., 2.5% at 300 K for SnSe. The significant anisotropy can be attributed to the different phonon group velocities along the three directions, as indicated by the phonon dispersion. For SnSe, we compare the *ab* initio results with measured $\kappa_L$ of both single- and poly-crystalline samples in Fig. 3 (a). For single crystals, we find that the calculated in-plane values are very close to the historically measured in-plane $\kappa_L$ (~ 1.9 W/mK at 300 K[24]). However, the calculated results are much higher than the recent experimental data[19] along the corresponding directions although the magnitude sequence $\kappa_L^b > \kappa_L^c > \kappa_L^a$ is consistent between the measured and *ab* initio results. Our results are also consistent with the finding in Ref. 31. It is also noted that the measured thermal conductivities for polycrystalline samples are much higher than the corresponding average values of single crystals in Ref. 19, which is unusual since the grain boundaries in polycrystalline samples will reduce the effective thermal conductivity. The polycrystal data lie between the calculated directional results and are lower than the average of the calculated directional values, which is expected for anisotropic materials with grains and possible defects[20]. The surprisingly lower $\kappa_L$ of measured single crystals might be caused by defects. The later analysis will show that optical phonons contribute ~ 60% to the total $\kappa_L$ along the three directions for SnSe. Assuming that the optical phonon contribution is completely suppressed by point-defect scattering, $\kappa_L$ at 300 K would be reduced to 0.3, 0.8 and 0.6 W/mK along the *a*, *b* and *c* directions, respectively. These values are close to the reported experimental data for single crystals and may explain the discrepancy between the recent measurements and calculations, and that between measured single- and poly-crystalline results. The calculated lattice thermal conductivities of SnS show similar agreements with the experimental results[24]. We notice that $\kappa_L^b$ and $\kappa_L^a$ in SnS are relatively higher than those in SnSe, resulting in ~ 8% higher average $\kappa_L$ in the former. This can be explained by the relatively larger phonon group velocities and longer relaxation times of optical modes between 2 and 3 THz, which will be discussed later.

**D. Phonon spectral analysis**

We further plot the cumulative directional lattice thermal conductivity with respect to the phonon MFP for SnSe and SnS at 300 and 750 K, as shown in Fig. 4. It is found that their phonon MFP distributions are similar, agreeing with the small $\kappa_L$ difference. For both materials, the thermal conductivities are dominated by phonons of short MFPs, e.g., at room temperature heat is mainly carried by phonons with MFPs ranging from 1 to 30 nm and the MFP values corresponding to 50% $\kappa_L$ accumulation are only ~ 5 nm, limiting the potential in reducing $\kappa_L$ by nanostructuring. At high temperatures, the phonon MFPs become even shorter, e.g., the MFP corresponding to the median $\kappa_L$ accumulation in SnSe reduces from 4.9 nm at 300 K to 1.9 nm at 750 K along the *b* axis. The phonon MFPs in SnSe and SnS are notably shorter than those in PbSe and PbTe (up to ~ 300 nm at 300 K), accounting for the lower $\kappa_L$ in the formers. Despite the anisotropic $\kappa_L$ along the *a*, *b* and *c* axes, the MFP distributions differ little along the three directions.

To understand the anisotropic $\kappa_L$ and the thermal transport difference between SnSe and SnS, we obtain the room temperature $\kappa_L$ contribution with respect to frequency along the *a*, *b* and *c* axes for both materials, as shown in Figs. 5 (a) and (b). One can find that the $\kappa_L$ contribution along the *a* axis is smaller than that along the in-plane directions for both materials and the anisotropic $\kappa_L$ is mainly caused by phonons with frequency < 3 and 4 THz for SnSe and SnS, respectively. Along the *b* and *c* axes, phonons with higher frequencies almost contribute equally to $\kappa_L$. To evaluate the influence of phonon group velocity and relaxation time on the anisotropy in $\kappa_L$, we define two quantities

$$\bar{v}_g^\alpha = \sqrt{\frac{\sum_{qs} C_V(\mathbf{q}s) v_g^\alpha(\mathbf{q}s) v_g^\alpha(\mathbf{q}s) \tau_{\mathbf{q}s}}{\sum_{qs} C_V(\mathbf{q}s) \tau_{\mathbf{q}s}}} \tag{2}$$

$$\bar{\tau}^\alpha = \frac{\sum_{qs} C_V(\mathbf{q}s) v_g^\alpha(\mathbf{q}s) v_g^\alpha(\mathbf{q}s) \tau_{\mathbf{q}s}}{\sum_{qs} C_V(\mathbf{q}s) v_g^\alpha(\mathbf{q}s) v_g^\alpha(\mathbf{q}s)} , \tag{3}$$

which represent the corresponding average values over the entire Brillouin zone along a particular direction. For SnSe, we obtain ($\bar{v}_g^a$ = 395 m/s, $\bar{v}_g^b$ = 627 m/s, $\bar{v}_g^c$ = 582 m/s) and ($\bar{\tau}^a$

= 3.4 ps, $\bar{\tau}^b$ = 3.2 ps, $\bar{\tau}^c$ = 3.1 ps) at 300 K, which clearly indicates the dominance of phonon group velocity on the anisotropic phonon transport although the group velocities of the long-wavelength acoustic phonons are close (Table 1). Similarly, the phonon group velocity determines the anisotropy in SnS, as indicated by the average values ($\bar{v}_g^a$ = 443 m/s, $\bar{v}_g^b$ = 712 m/s, $\bar{v}_g^c$ = 598 m/s) and ($\bar{\tau}^a$ = 1.9 ps, $\bar{\tau}^b$ = 2.2 ps, $\bar{\tau}^c$ = 1.8 ps). A larger difference can be found in $\bar{v}_g^\alpha$ and $\bar{\tau}^\alpha$ between $b$ and $c$ directions for SnS, accounting for the $\kappa_L$ difference along these two directions.

Based on the acoustic cutoff frequencies for SnSe (1.6 THz) and SnS (1.9 THz), the optical phonons can be found to contribute significantly to $\kappa_L$ along the three directions. In Figs. 6 (a) and (b), we further show the accurate $\kappa_L$ contributions from different phonon branches along the $b$ axis. The optical phonons in SnSe and SnS contribute respectively ~ 61% and 70% to the total $\kappa_L$ from 300 to 750 K, which are surprisingly large considering that 20% of the room temperature $\kappa_L$ comes from optical phonons in PbSe and PbTe belonging to another group of IV-VI compounds. Such a behavior results from the long MFPs and large DOS of optical phonon modes. Due to the large group velocities and significant relaxation times (see Fig. 6 (d)), the MFPs of many optical phonon modes are comparable with those of acoustic phonons and can be longer than 10 nm along the $b$ axis. In regards to acoustic phonons, the $\kappa_L$ contribution of the first transverse branch (TA1) is almost double that of the other two acoustic branches. It is also found that the relatively larger $\kappa_L$ in SnS is due to the contribution of optical phonons while the $\kappa_L$ contributed by acoustic phonons in SnS is slightly smaller than that in SnSe. We find that this difference is mainly caused by relaxation times, as shown in Figs. 6 (c) and (d). One can find that SnSe has more acoustic modes with relaxation times longer than 10 ps while the relaxation times of optical modes in SnS are much longer than those in SnSe between 2 and 3 THz, accounting for the corresponding large difference of $\kappa_L$ contributions between SnSe and SnS. Due to the same reason, a similar difference is observed in the MFPs between SnSe and SnS because there is only a small difference between their phonon group velocities. For example, SnSe has more acoustic phonon modes with MFPs longer than 30 nm compared with SnS. That is why the normalized cumulative $\kappa_L$ of SnSe approaches 1 at relatively longer phonon MFPs.

The relaxation times of SnSe and SnS are much shorter in comparison with those of many other bulk materials (e.g., ~ 100 ps for PbTe[42] and ~ 200 ps for CoSb$_3$[40] at $f$ = 0.4 THz and $T$ = 300 K), indicating very high anharmonicity. To further evaluate the anharmonicity, we calculated the Grüneisen parameters $\gamma_G = \sum_{\mathbf{q}s} \gamma(\mathbf{q}s) C_V(\mathbf{q}s) / \sum_{\mathbf{q}s} C_V(\mathbf{q}s)$ ($\gamma(\mathbf{q}s)$ is mode Grüneisen parameter) averaged over all phonon modes, leading to 2.12, 1.55 and 1.66 for SnSe and 2.17, 1.44 and 1.55 for SnS along the *a*, *b* and *c* axes, respectively. These values are comparable to the Grüneisen parameters of 2.05 for AgSbTe$_2$ ($\kappa_L$ = 0.68 W/mK)[53] and 1.45 for PbTe ($\kappa_L$ = 2.4 W/mK)[54] but lower than previously reported values (4.1, 2.1 and 2.3) for SnSe averaged over only the acoustic modes[19]. For both SnSe and SnS, the Grüneisen parameters along the *a* axis are significantly larger than those along *b* and *c* axes, which is related to the weak bonding between layers.

## III. PREDICTION OF ELECTRICAL PROPERTIES

### A. Calculation method

The electronic properties of SnSe and SnS were calculated using the BoltzTraP code[43] within the framework of the Boltzmann transport theory combined with the relaxation times obtained from the single parabolic band (SPB) model[44]. The implementation of the BoltzTrap is based on the full band structures in the entire Brillouin zone, which were calculated using the *ab initio* many-body perturbation theory within the GW approximation, as implemented in Ref. 55. This approach has been successfully applied in the predictions of electronic band structures of various semiconductors, showing good agreements with experimental results[23,55]. We started from the ground state calculations using the experimental lattice constants and atomic positions for SnSe ($a$ = 11.58 Å, $b$ = 4.22 Å, $c$ = 4.40 Å, *Pnma* at 600 K)[56] and SnS ($a$ = 11.20 Å, $b$ = 3.987 Å, $c$ = 4.334 Å, *Pnma* at 295 K)[57], with the Perdew-Burke-Ernzerhof (PBE)[58] form of generalized gradient approximation (GGA) as the exchange-correlation potential, which is typically used for predicting electronic band structures[23,55]. A 4×12×12 **k**-point sampling grid and a cut-off energy of 400 eV were adopted on both materials for convergence. We then updated the eigenvalues in the Green's function and the dielectric matrix by iteration until self-consistency was achieved. The band structures were then obtained using the Wannier interpolation scheme[59].

The constant relaxation time assumption adopted by BoltzTrap has limited the prediction of *ZT*. A possible routine is to use a constant relaxation time extracted from the experiments. However, the resulting prediction accuracy is questionable because the relaxation time varies with the temperature and carrier concentration, as indicated in previous investigations[20]. In this study, the relaxation times were obtained from the SPB model, which has been successfully applied to many materials[44,60,61]. In the SPB model, all transport coefficients can be obtained by solving the Boltzmann transport equation using an energy ($E$) dependent relaxation time

$$\tau = \tau_0 E^r \ . \tag{4}$$

Here, $r$ is related to the charge scattering mechanism. We only considered the electron scattering by acoustic phonons, which is dominant at moderate temperatures, and $r$ = -1/2 is therefore used for subsequent calculations. The relaxation time limited by the acoustic phonon scattering is given by

$$\tau = \frac{2^{1/2}\pi\hbar^4 \rho v_l^2}{3E_d^2 (m^*)^{3/2} (k_B T)^{3/2}} \frac{F_0(\eta)}{F_{1/2}(\eta)} \,, \tag{5}$$

where $\hbar$ is the reduced Planck constant, $\rho$ is the mass density, $v_l$ is the longitudinal sound velocity, $E_d$ is the deformation potential, $m^*$ is the single valley density-of-states effective mass, $k_B$ is the Boltzmann constant, $T$ is the temperature and $\eta$ ($=E_F/k_B T$) is the reduced chemical potential. The Fermi integral $F_x(\eta)$ is represented as

$$F_x(\eta) = \int_0^\infty \frac{E^x}{1+\exp(E-\eta)} dE \,. \tag{6}$$

The electron relaxation time $\tau$ is assumed to be direction-independent, which has been shown to be reasonable even in substantially anisotropic structures[62]. Therefore, the directional carrier mobility $\mu_\alpha$ can be determined by $\mu_\alpha = \tau e / m_\alpha^*$, where $e$ is the electron charge and $m_\alpha^*$ is the effective mass along the direction denoted by $\alpha$.

The carrier concentration $n$ is determined by

$$n = \frac{(2m_d^* k_B T)^{3/2}}{2\pi^2 \hbar^3} F_{1/2}(\eta) \,, \tag{7}$$

where $m_d^*$ is the density-of-states effective mass and is related to the valley degeneracy $N_V$ by $m_d^* = N_V^{2/3} m^*$.

The electrical conductivity $\sigma_\alpha$ can then be obtained from $\sigma_\alpha = ne\mu_\alpha$. The Hall carrier mobility $\mu_H$ is related to the electrical conductivity by $\mu_H^\alpha = \sigma_\alpha / en_H$, where $n_H$ is inversely proportional to the Hall coefficient. In the following discussions, we use $n_H$ for the carrier concentration and $\mu_H$ for the carrier mobility in order to compare with the experimental results.

The electronic thermal conductivity $\kappa_e^\alpha$ is calculated according to the Wiedemann-Franz law $\kappa_e^\alpha = L\sigma_\alpha T$, where the Lorenz number $L$ is determined from the BTE, i.e., extracting the ratio of the electronic thermal conductivity to the electrical conductivity multiplied by temperature as implemented in the BoltzTrap.

To accurately predict the relaxation time, we calculated all the necessary parameters from first-principles. The effective masses $m^*_{k,l} = \hbar^2/[\partial^2 E / \partial k_k \partial k_l]$ of holes and electrons along the three directions were calculated near the band extrema. The deformation potential (DP) for the $i$-th band is defined as $E_d^i = \partial E / (\partial a / a_0)$ ($a_0$ is the equilibrium lattice constant). In this work, the $E_d$ for holes and electrons are calculated based on the energy change of the valence band maximum (VBM) and conduction band minimum (CBM), respectively. We chose a series of lattice constants (0.99$a_0$, 0.995$a_0$, $a_0$, 1.005$a_0$ and 1.01$a_0$) for the calculation of the DPs. The absolute DPs were obtained by choosing the energy level of the deep core state as the reference, which can be assumed to be insensitive to the slight lattice deformations, as suggested by Wei and Zunger[63]. The longitudinal sound velocities were determined according to the phonon dispersions near the Gamma point. All the obtained parameters are listed in Table I.

TABLE I: Parameters obtained from first-principles used for the SPB model.

| Material | Carrier | $m^*$ ($m_0$) | | | $m_d^*$ ($m_0$) | $E_d$ (eV) | | | $v_l$ (m/s) | | |
|---|---|---|---|---|---|---|---|---|---|---|---|
| | | a | b | c | | a | b | c | a | b | c |
| SnSe | VBM | 0.78 | 0.18 | 0.30 | 0.55 | 14.1 | 16.4 | 15.8 | 3356 | 3493 | 3267 |
| | VBM1 | 0.46 | 0.16 | 0.24 | 0.41 | | | | | | |
| | CBM | 0.50 | 0.12 | 0.16 | 0.34 | 12.9 | 10.3 | 13.2 | | | |
| | CBM1 | 0.08 | 1.18 | 1.29 | 0.50 | | | | | | |

| | | | | | | | | |
|---|---|---|---|---|---|---|---|---|
| | VBM | 1.35 | 0.21 | 0.36 | 0.74 | 14.9 | 21.9 | 19.1 |
| | VBM1 | 0.43 | 0.19 | 0.30 | 0.46 | | | |
| SnS | | | | | | | 3750 | 3823 | 3625 |
| | CBM | 0.51 | 0.15 | 0.20 | 0.39 | 14.3 | 11.0 | 14.6 |
| | CBM1 | 0.09 | 1.02 | 1.14 | 0.47 | | | |

## B. Electronic band structure

Figure 7 shows the GW quasiparticle band structures of SnSe and SnS along some high symmetry directions, which are found to be very similar, with a VBM [at (0.00, 0.00, 0.36) for SnSe and (0.00, 0.00, 0.43) for SnS] along the Γ→Z direction and a CBM [at (0.00, 0.32, 0.00) for both materials] along the Γ→Y direction. The quasiparticle energy shifts are nearly uniform with respect to the initial GGA calculations, resulting in band gaps of 0.78 eV for SnSe and 1.05 eV for SnS, respectively. These values are in good agreements with the band gaps obtained from optical measurements (0.86-0.948 eV for SnSe[19,64,65] and 0.9-1.142 eV for SnS[64,66]) and previous GW predictions[23,29]. Remarkable anisotropy is found for the effective masses at VBM and CBM, which agree with the experimental data[62] and previous *ab initio* results[23]. For instance, the calculated effective masses of hole along the *a, b* and *c* axes are 1.35 $m_0$, 0.21 $m_0$ and 0.36 $m_0$, agreeing well with the cross-plane (1.0 $m_0$) and in-plane values (0.2 $m_0$) obtained by the infrared measurements on SnS at room temperature[62]. Along most directions, the effective masses of hole and electron for SnSe are relatively smaller than those for SnS, which might lead to a large difference in carrier mobility between them. For both materials, the effective mass of hole is typically larger than that of electron along the same direction, especially along the *a* direction in SnS. One can also note a local conduction band minimum (CBM1) at the Γ point (0.00, 0.00, 0.00) and a local valence band maximum (VBM1) [at (0.00, 0.31, 0.00) for SnSe and (0.00, 0.30, 0.00) for SnS] along the Γ→Y direction in both band structures. SnSe has a CBM1 that is 0.05 eV higher than the global CBM and a VBM1 that is 0.15 eV lower than the global VBM while SnS exhibits correspondingly a 0.22 eV higher CBM1 and a 0.20 eV lower VBM1. It should be noted that the relative positions between these band edges may change with the temperature, as shown in Ref. 30. As the doping level increases, these energy bands will also contribute to the electronic properties. For example, the chemical potential reaches the CBM1 for a carrier concentration of ~ $2 \times 10^{19}$ cm$^{-3}$ at 300 K in the *n*-type SnSe. Further, the effective masses of these valleys are quite different, as shown in Table I. Specifically, the effective masses of electrons become significantly smaller along the *a* axis but much larger along the *b* and *c* axes at the CBM1. These changes will dramatically affect the density of states, carrier mobilities, Seebeck coefficients and other related properties. Therefore, we considered the contributions of both the band valley and the secondary valleys in the calculations of electronic properties for both *n*- and *p*-type materials. Importantly, the relaxation times for different valleys are calculated based on corresponding effective masses. We found that such a treatment is valid for the carrier concentration up to $1 \times 10^{21}$ cm$^{-3}$ from 300 to 750 K. It should be noted that the chemical potential goes beyond the two band valleys for significantly higher $n_H$ (e.g., $1 \times 10^{22}$ cm$^{-3}$) and therefore more bands should be considered in the calculations.

**C. Electrical conductivity**

Despite the improved accuracy of the GW scheme, the band gaps of SnSe and SnS are slightly underestimated. For better comparison with experimental data, we calculated the electrical properties by shifting the band gaps of SnSe and SnS to the experimental values 0.86 eV and 1.11 eV, respectively, while keeping the band structure shape. The electrical conductivities of SnSe and SnS are shown in Fig. 8. Note that in this work the Hall carrier concentration $n_H$ is the value corresponding to the given temperature since it is temperature-dependent, e.g., $n_H$ of a $p$-type SnSe sample can increase from $2 \times 10^{17}$ cm$^{-3}$ to $5 \times 10^{18}$ cm$^{-3}$ when the temperature changes from 300 to 750 K[20]. For semiconductors, the Hall coefficient $R_H = (p\mu_h^2 - n\mu_e^2)/[e(p\mu_h + n\mu_e)^2]$ is related to the carrier concentrations ($n$ for electron and $p$ for hole) and mobilities ($\mu_e$ for electron and $\mu_h$ for hole), which all vary with temperature[67]. From Fig. 8 (a), one can find that $\sigma$ of the $p$-type SnSe is highly anisotropic at 300 and 750 K, characterized by the largest one along the $b$ axis, the smallest one along the $a$ axis and that along the $c$ axis is in between. For example, $\sigma_b$ is around four times larger than $\sigma_a$ for $n_H = 1.0 \times 10^{19}$ cm$^{-3}$ at 300 K. A comparison between the measured (polycrystals) and calculated results is shown for the $p$-type SnSe at 750 K. The variation tendency of $\sigma$ with respect to the carrier concentration is found to be well predicted. Also, $\sigma$ of the $n$-type SnSe (see Fig. 8 (b)) exhibits similar anisotropy for a low carrier concentration at 300 and 750 K. However, due to different growing rates, $\sigma_a$ gradually approaches $\sigma_b$ and $\sigma_c$ at higher $n_H$. This behavior results from the increasing relative contribution of electrons in the secondary band valley when the chemical potential approaches the CBM1. The secondary band valley favors both $n$ due to the larger density of states effective mass and $\mu_a$ due to the significantly smaller $m_a^*$. One can also find that the anisotropy of $\sigma$ at 750 K is smaller than that at 300 K at low carrier concentrations. This can be attributed to the larger relative contribution of the second valley at higher temperatures considering that the electronic transport properties are dominated by the states with an energy within $3k_BT$ around the chemical potential. Typically, $\sigma$ of $n$-type SnSe is significantly larger than that of the $p$-type one at the same carrier concentration, which is due to the higher electron mobilities resulting from the smaller effective masses (see Table I). Similar anisotropy is found for the electrical conductivity in SnS. Because the energy difference between CBM and CBM1 in SnS is much larger than that in SnSe, the enhancement effect is not as significant as that in SnSe. Due to the relatively larger effective masses, electrical conductivities in SnS are significantly lower than those in SnSe.

**D. Hall Carrier mobility**

We further calculated the mobilities of holes and electrons with respect to the carrier concentration and compared them with the experimental results. As shown in Fig. 9 (a), the hole mobility of SnSe shows high anisotropy characterized by $\mu_H^b > \mu_H^c > \mu_H^a$, which is consistent with the relative values of the hole effective masses. This tendency is also consistent with the experimental data[19,20] at both 300 and 750 K and the calculated directional values reasonably agree with the measured results for single crystals, verifying the accuracy of this approach. It is also noted that the predicted average values are relatively larger than most measured results for polycrystalline samples. A possible reason is the existence of secondary phases (such as $AgSnSe_2$ for Ag doping) and defects in the samples, which can significantly lower the carrier mobility. The electron mobility (see Fig. 9 (b)) shows similar anisotropy along the three directions at 300 K when the carrier concentration is low, which is consistent with the effective masses of electrons. Also, $\mu_H^a$ gradually approaches $\mu_H^b$ and $\mu_H^c$ as $n_H$ increases, which is consistent with the variation of $\sigma$. The increasing relative contribution of electrons in the secondary valley slows down the decrease of the electron mobilities along the *a* axis, as clearly indicated by the slopes between $3 \times 10^{19}$ cm$^{-3}$ and $2 \times 10^{20}$ cm$^{-3}$ at 300 K. At a given chemical potential, $\mu_H^a$ of electrons in the secondary valley is around two times that of the first valley. From Figs. 9 (c) and (d), one can find that the mobilities of holes and electrons in SnS exhibit similar variations. The calculated $\mu_H^b$ and $\mu_H^c$ show good agreements with the in-plane experimental data[62,68] for the *p*-type single crystalline SnS. Mainly due to the relatively smaller effective masses in the first valence and conduction band valleys, SnSe shows significantly larger mobilities for both hole and electron compared with SnS. In both materials, the mobilities of electrons are significantly larger than those of holes because of the smaller effective masses in the first conduction band valleys and deformation potentials.

**E. Seebeck coefficient**

Figure 10 shows the calculated Seebeck coefficients as a function of carrier concentration for *p*- and *n*-type SnSe and SnS at 300 and 750 K. The average Seebeck coefficients are calculated by $S_{avg} = (S_a \sigma_a + S_b \sigma_b + S_c \sigma_c)/(\sigma_a + \sigma_b + \sigma_c)$, which therefore represent those of polycrystals[29]. As shown in Fig. 10 (a), the Seebeck coefficients of *p*-type SnSe along the three directions are found to be close to each other at 300 and 750 K, agreeing with the experimental data for single crystals[19]. That's why the measured Seebeck coefficients of the single- and poly- crystalline *p*-type SnSe are comparable[19,20]. At both 300 and 750 K, the average values reasonably agree with the experimental data for the *p*-type SnSe but are slightly overestimated. At 300 K, the *p*-type SnSe shows significantly larger magnitudes of Seebeck coefficients along the three directions than the *n*-type counterpart for the carrier concentration below $2 \times 10^{19}$ cm$^{-3}$, which can be ascribed to the larger density of states effective mass of hole[69]. With the increase of the carrier concentration, a special behavior arises for the $S_a$ in the *n*-type SnSe, which becomes almost a constant from $2 \times 10^{19}$ cm$^{-3}$ to $1 \times 10^{20}$ cm$^{-3}$ and thus results in notably larger magnitudes of the Seebeck coefficient with respect to the in-plane ones and those of the *p*-type counterpart. This behavior is due to the enhancement in $S_a$ that can be explained by the Mott formula

$$S = \frac{\pi^2}{3} \frac{k_B}{e} k_B T \left\{ \frac{1}{n} \frac{dn(E)}{dE} + \frac{1}{\mu} \frac{d\mu(E)}{dE} \right\}_{E=E_F}, \quad (8)$$

with $n(E) = D(E)f(E)$, where $n(E)$, $D(E)$, $f(E)$ and $\mu(E)$ represent the carrier density, density of states, Fermi-Dirac distribution function and carrier mobility at the energy level $E$, respectively[4,70]. As the chemical potential approaches the second conduction band valley, the energy-dependence of $n(E)$ and $\mu_a(E)$ is increased due to the larger density of states effective mass and the significantly smaller $m_a^*$, respectively. This enhancement effect also extends to high temperatures. For example, the Seebeck coefficient ratio $S_a/S_b$ increases from 1.22 for $n_H$ of $1 \times 10^{18}$ cm$^{-3}$ to 1.41 for $n_H$ of $5 \times 10^{19}$ cm$^{-3}$ at 750 K for the same *n*-type SnSe sample. Similar behavior is observed in the Seebeck coefficients of SnS. For SnS, the enhancement effect arises at higher $n_H$ due to the higher energy position of the CBM1, e.g., the chemical potential reaches the CBM1 at $n_H$ of ~ $2 \times 10^{20}$ cm$^{-3}$ at 300 K. Typically, both *p*- and *n*-type SnS show larger Seebeck values with respect to the SnSe counterparts within the considered carrier concentration range, which is consistent with the larger band gap in SnS. Indeed, we also obtained the chemical potential dependence of the Seebeck coefficients and found that the ratio of peak values between SnSe and SnS are very close to the ratio of corresponding band gaps, a typical behavior for thermoelectric semiconductors[71]. For both SnSe and SnS, the temperature dependence of the Seebeck coefficient is significant.

**F. Figure of merit**

The calculated electronic thermal conductivity shows similar anisotropy along the three directions. Typically, the electronic contribution to the thermal conductivity is negligible for lightly-doped samples, e.g., the average electronic thermal conductivities are smaller than 0.1 W

/mK in *p*-type SnSe at 750 K when the carrier concentration is lower than $3 \times 10^{19}$ cm$^{-3}$. The effects of dopants on the lattice thermal conductivity are considered based on Ag and In, which are promising candidates to achieve *p*-type[20,21] and *n*-type[72] SnSe (SnS), respectively.

After obtaining the carrier concentration dependence of all these properties, we were able to calculate *ZT*s for both SnSe and SnS at 750 K, as shown in Fig. 11. To calculate the average *ZT*, the power factor (PF) is averaged over the three directions by $PF_{avg} = (S_a^2 \sigma_a + S_b^2 \sigma_b + S_c^2 \sigma_c)/3$, corresponding to that of polycrystalline materials of *large grains*. Because doping generally benefits the electrical conductivity while deteriorates the Seebeck coefficient, *ZT* will first increase and then decrease with the carrier concentration, therefore leading to an optimal value. Figure 11 (a) shows *ZT*s of the *p*-type SnSe at 750 K, in comparison with the experimental results for single- and poly-crystalline samples. One can find a significant anisotropy for *ZT* that is largest along the *b* axis and smallest along the *a* axis in the *p*-type SnSe for $n_H$ lower than $1 \times 10^{20}$ cm$^{-3}$, which is consistent with the experimental results[20]. The optimal *ZT*s arise at similar carrier concentrations along the three directions, e.g., $\sim 6 \times 10^{19}$ cm$^{-3}$ for the *p*-type SnSe. The experimental *ZT*s for single crystals are significantly larger than the predicted values, which is mainly due to the significantly lower measured thermal conductivity, as discussed above. The polycrystalline results are close to the calculated average values when $n_H$ is below $2 \times 10^{19}$ cm$^{-3}$ but decrease significantly at higher $n_H$. Two possible reasons may account for the discrepancy. First, the volume fraction of metallic secondary phases (e.g., AgSnSe$_2$) may increase at higher $n_H$, which will enhance the thermal conductivity while reducing the Seebeck coefficient and carrier mobility[20]. Second, the polycrystalline data may deviate from the average values because the measurement is sensitive to the hot press direction. The maximum average *ZT* for the *p*-type SnSe at 750 K is 0.8 while the peak value along the *b* axis can reach 1.0.

The thermoelectric performance of *n*-type SnSe is found to be more promising, as shown in Fig. 11 (b). The optimal *ZT*s along all directions in the *n*-type SnSe are remarkably larger than those in the *p*-type one. Different from the *p*-type one, the *ZT* values along the *a* axis are highest while those along the *b* axis are lowest in the *n*-type SnSe, which are due to the enhancement and depression of the power factor, respectively. Without considering the enhancements in $S_a$ (evaluated by $S_a = (S_b + S_c)/2$) and $\sigma_a$ (assuming the effective mass remains at the CBM1), the optimal *ZT* (See "*a*-1" in Fig. 11 (b)) along the *a* axis is only 1.3. This peak value (See "*a*-2" in Fig. 11 (b)) reaches 1.5 if only $\sigma_a$ is enhanced. Further enhancement in $S_a$ leads to a much larger optimal *ZT* of 2.7 in the *n*-type SnSe. Such a high *ZT* suggests great potential performance in the single crystalline SnSe. Furthermore, the maximum average *ZT* in the *n*-type SnSe can be as high as 1.5, which is competitive with the reported highest values for *p*-type PbTe (1.5 at 773 K[4]) and PbSe (1.2 at 850 K[8]). For both types, the optimal *ZT*s of SnSe arise within the carrier concentration ranging from $1 \times 10^{19}$ cm$^{-3}$ to $1 \times 10^{20}$ cm$^{-3}$ and the *n*-type one requires a lighter doping to reach the maximum *ZT*. For SnS, similar variations of *ZT*s are observed for both *p*- and *n*-type dopings. Compared with SnSe, SnS possesses lower optimal *ZT*s along the three directions, which is mainly due to the lower carrier mobilities. Although the highest *ZT* is only 0.6 along the *b* axis in the *p*-type SnS, that in the *n*-type one can reach 1.5 along the *a* axis and 1.0 for the average value, which are still competitive.

## IV. CONCLUSIONS

We performed comprehensive first-principles calculations to predict the thermal and electrical transport properties for *p*- and *n*-type SnSe and SnS. The phonon transport details show that SnSe and SnS exhibit similar thermal properties while the average $\kappa_L$ of SnS is ~ 8% higher than that of SnSe from 300 to 750 K. Remarkable anisotropy ($\kappa_L^b > \kappa_L^c > \kappa_L^a$) in the lattice thermal conductivity is found for both materials, which is mainly due to the anisotropic group velocity resulting from the layered lattice structure. The cross-plane $\kappa_L$ is significantly lower than the in-plane ones. The phonon mode relaxation times indicate high anharmonicity in both materials, accounting for the extremely low $\kappa_L$. The very short MFPs of dominant phonon modes (1-30 nm at 300 K) suggest that nanostructuring would be difficult to further decrease the $\kappa_L$. It is also found that the relative $\kappa_L$ contribution (~ 65%) of optical phonons is significantly larger than that of acoustic phonons because the MFPs of many optical phonons are comparable with those of acoustic phonons. As for the absolute $\kappa_L$ contribution, that of acoustic (optical) phonons for SnSe is larger (smaller) than that of SnS mainly due to the relaxation time difference.

Similarly, both SnSe and SnS show significantly anisotropic electrical conductivities mainly due to the different effective masses along the *a*, *b* and *c* directions ($m_a^* > m_c^* > m_b^*$ for both holes and electrons at the VBM and CBM, respectively). The anisotropy for the Seebeck coefficient is much smaller. For the *p*-type materials, the anisotropy $\sigma_b > \sigma_c > \sigma_a$ remains when the carrier concentration spreads from $1 \times 10^{18}$ cm$^{-3}$ to $1 \times 10^{21}$ cm$^{-3}$ at 300 and 750 K. Consequently, the *ZT*s reach highest optimal values along the *b* axis and lowest peaks along the *a* axis. Interestingly, along the *a* direction in *n*-type materials, the electron mobilities in the secondary band valley are significantly larger than those in the first valley. This significantly benefits the electrical conductivity and the Seebeck coefficient as the chemical potential gradually approaches the secondary conduction band valley, leading to the highest *ZT*s in combination with the lowest lattice thermal conductivities along the *a* direction. This finding indicates a promising routine to develop so called "phonon-glass electron-crystal" thermoelectric materials by tuning the band structures. SnSe has higher hole and electron mobilities compared with SnS, which is mainly caused by the smaller effective masses of SnSe in the first valence and conduction band valleys. Meanwhile, the lattice conductivities of SnSe are relatively lower. Therefore, SnSe exhibits larger optimal *ZT*s for both *p*- and *n*-type dopings. For both materials, the smaller effective masses of electrons lead to higher electron mobilities than hole mobilities along the three directions, leading to higher *ZT*s in *n*-type ones. Compared with the highest *ZT*s of 1.0 for SnSe and 0.6 for SnS along the *b* axis for the *p*-type doping, those for the *n*-type doping can reach 2.7 for SnSe and 1.5 for SnS along the *a* axis at 750 K. Furthermore, the optimal average *ZT*s are as high as 1.5 and 1.0 for the *n*-type SnSe and SnS at 750 K, respectively. Therefore, both single- and poly-crystalline *n*-type SnSe and SnS are very promising for large-scale thermoelectric applications.

In light of these findings, we conclude that there is still considerable potential to improve the thermoelectric performance of SnSe and SnS, especially for the *n*-type materials. The two major strategies to enhance *ZT*, i.e., reducing the lattice thermal conductivity and favoring the power factor, can be further explored based on the phonon and electron transport details obtained in this work.

**ACKNOWLEDGMENTS**

We are thankful for the financial support from the Hong Kong General Research Fund (Grant Nos. 613413 and 623212) and the National Natural Science Foundation of China (Grant No. 51376154).

* mebhuang@ust.hk

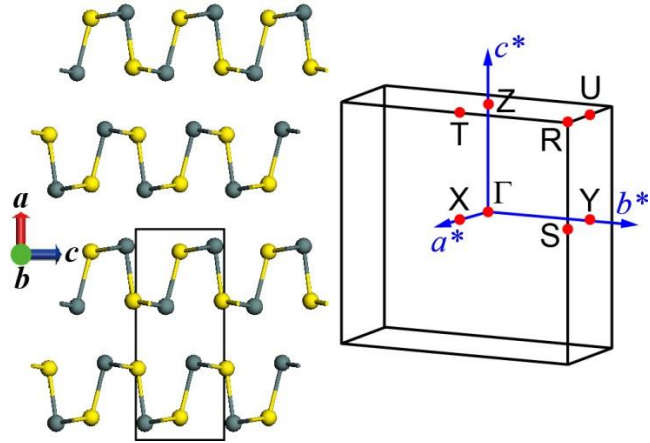

Figure 1. Crystal structure (Left) of the low temperature (*Pnma*) phase for SnSe (SnS) (grey: Sn atoms, yellow: Se (S) atoms) and the corresponding first Brillouin zone (Right) with high-symmetry points (red). The black border indicates a conventional unit cell consisting of 8 atoms.

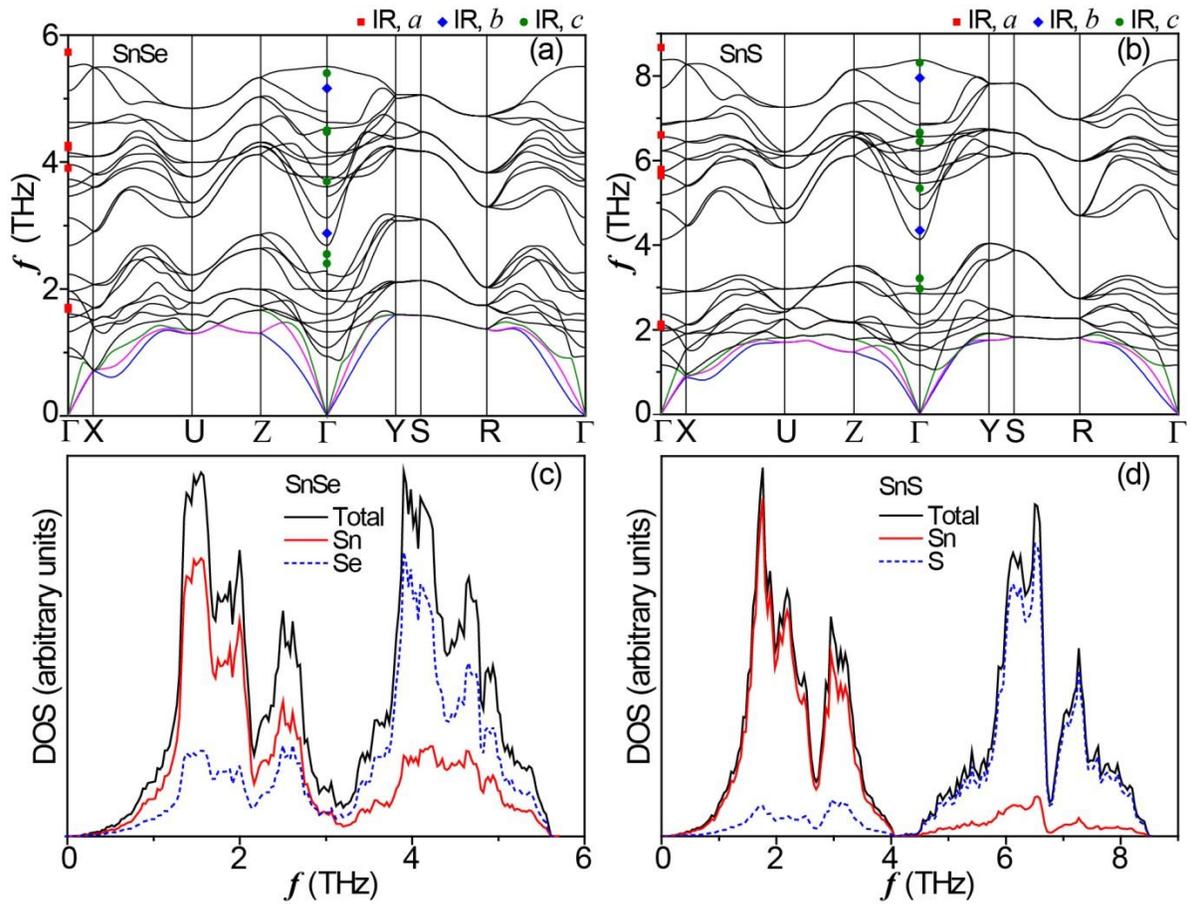

Figure 2. Calculated phonon dispersions along different high-symmetry paths (refer to the Brillouin zone in Fig. 1) for SnSe (a) and SnS (b). Experimental results[51] are denoted by filled symbols. The total and partial phonon densities of states for SnSe (c) and SnS (d) are also shown.

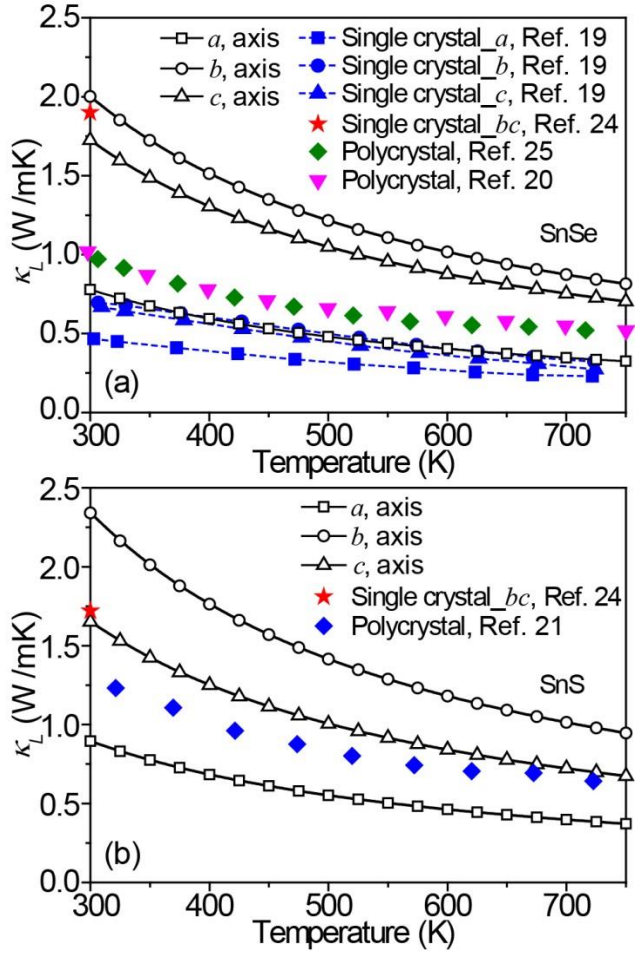

Figure 3. Calculated $\kappa_L$ (denoted by the open symbol and solid line) of SnSe (a) and SnS (b) along the *a*, *b* and *c* axes from 300 to 750 K, in comparison with the experimental data[19-21,24,25].

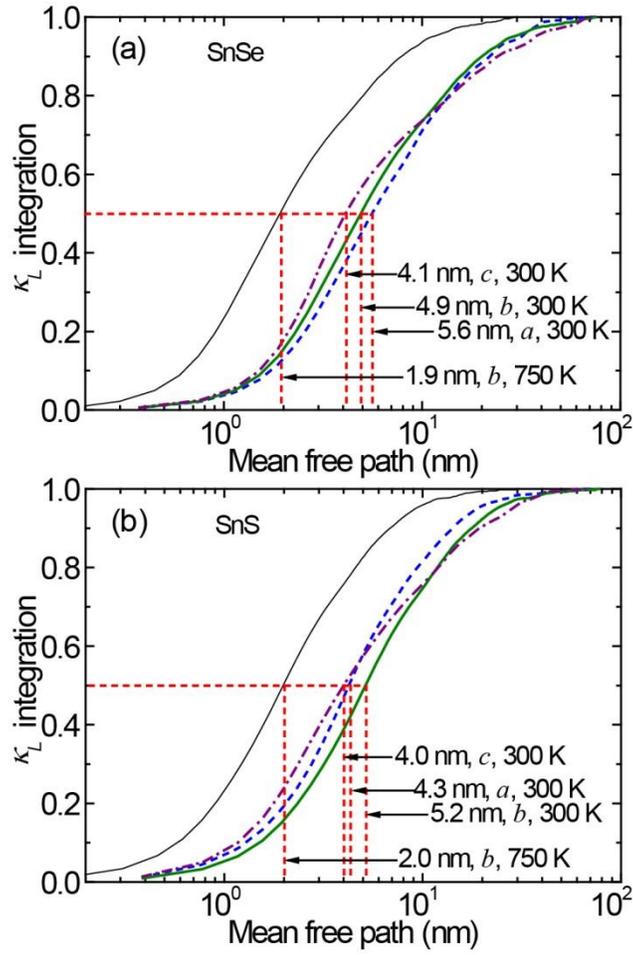

Figure 4. Normalized $\kappa_L$ integration for SnSe (a) and SnS (b) with respect to the phonon mean free path along different directions.

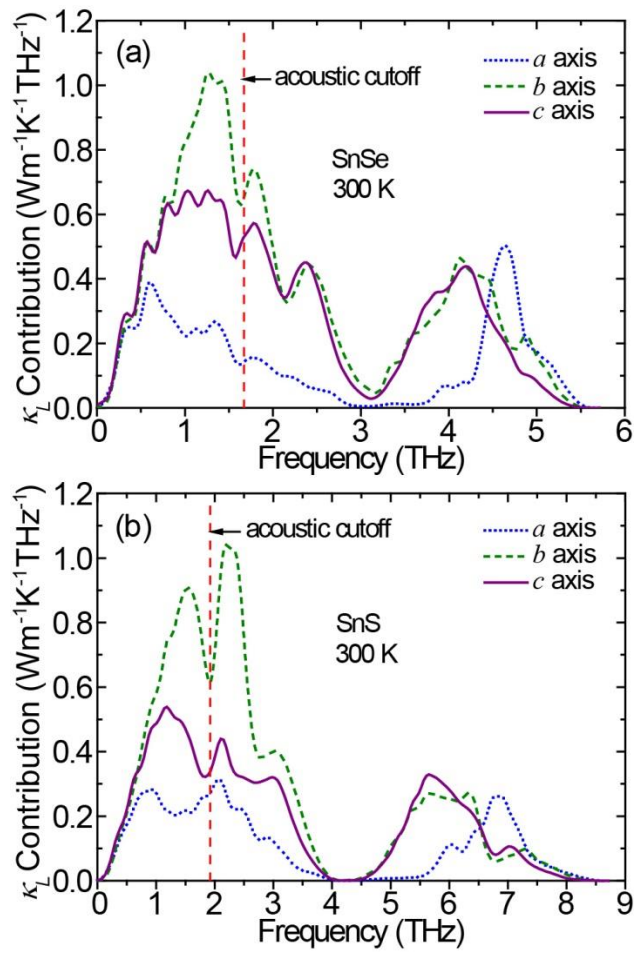

Figure 5. $\kappa_L$ contribution for SnSe (a) and SnS (b) with respect to frequency along the *a*, *b* and *c* axes at 300 K.

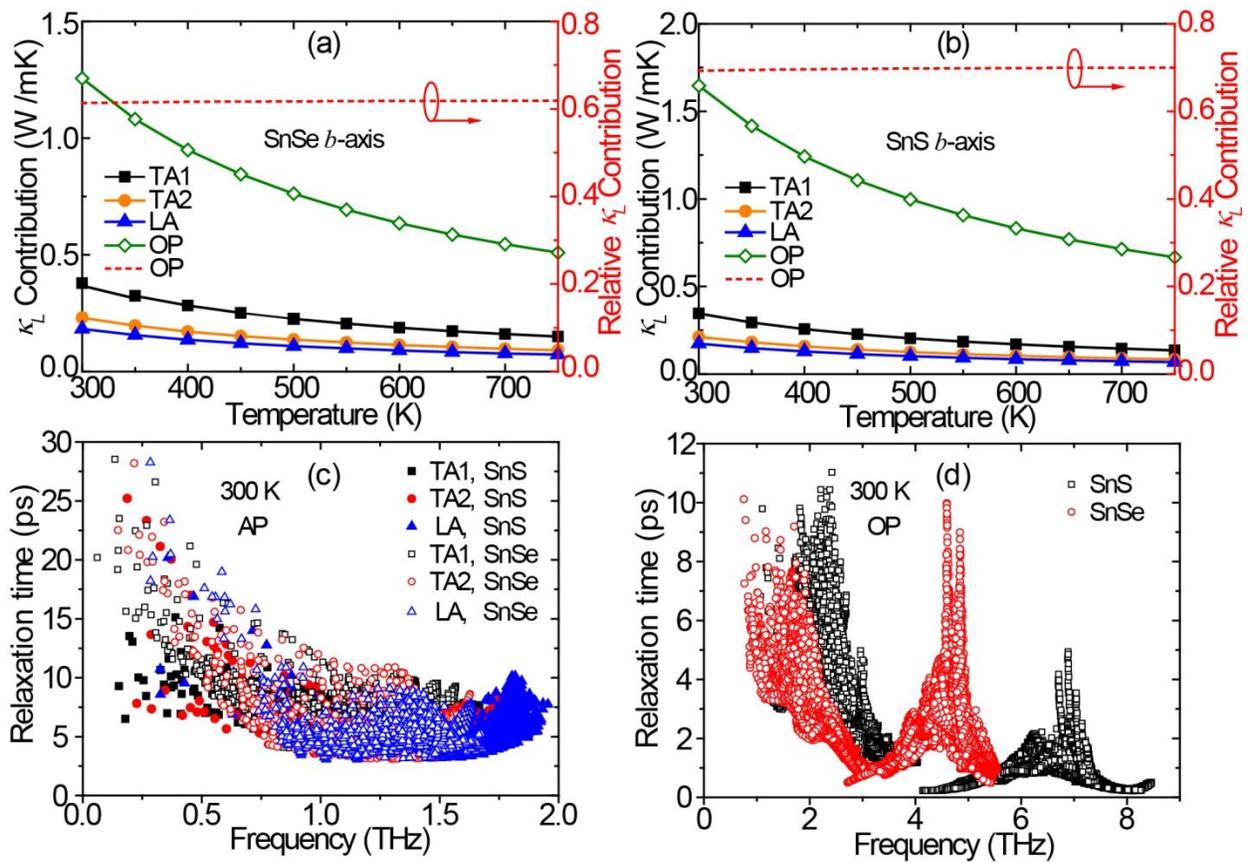

Figure 6. $\kappa_L$ contribution of acoustic (transverse: TA1, TA2 and longitudinal: LA) and optical (OP) phonon branches along the *b* axis from 300 to 750 K for SnSe (a) and SnS (b). (c) and (d) show the frequency-dependent phonon relaxation times of acoustic (AP) and optical phonons for SnSe and SnS, respectively, at 300 K.

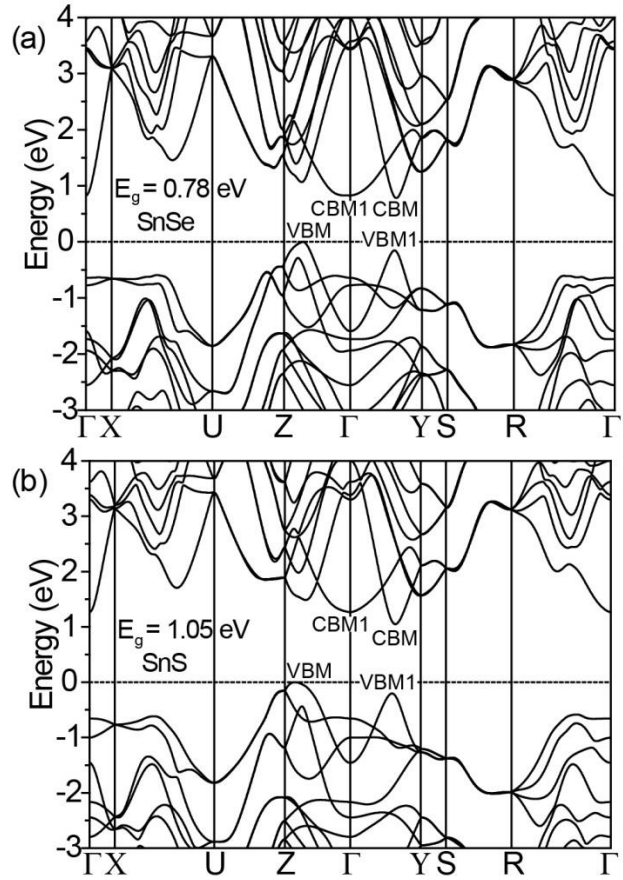

Figure 7. Electronic band structures computed for SnSe (a) and SnS (b). The valence band maximum, conduction band minimum, local valence band maximum and local conduction band minimum are marked as VBM, CBM, VBM1 and CBM1, respectively.

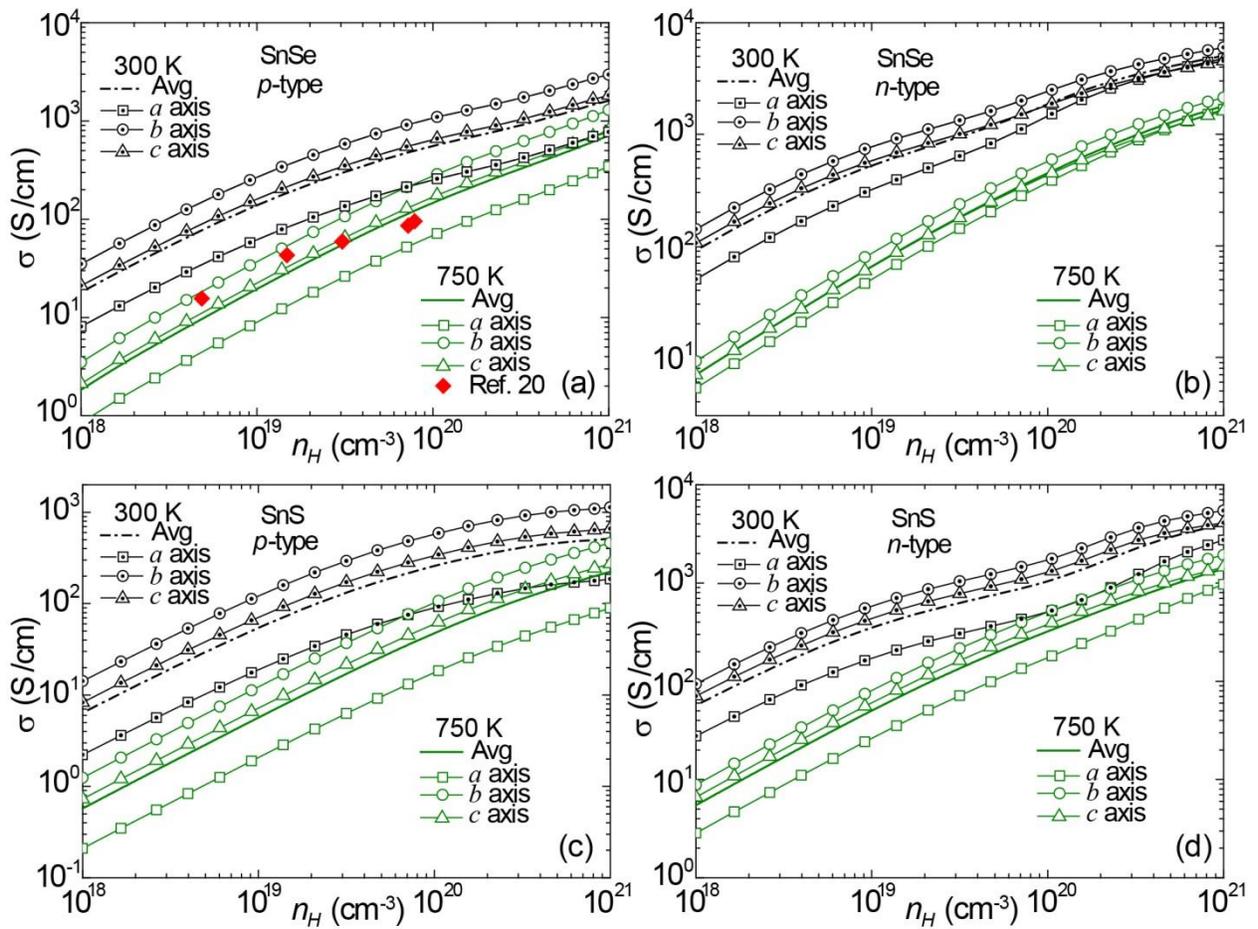

Figure 8. Electrical conductivities along the *a*, *b*, *c* axes and the average ones calculated at 300 and 750 K for *p*- and *n*-type SnSe [(a) and (b)] and SnS [(c) and (d)], in comparison with the experimental results for polycrystalline samples[20].

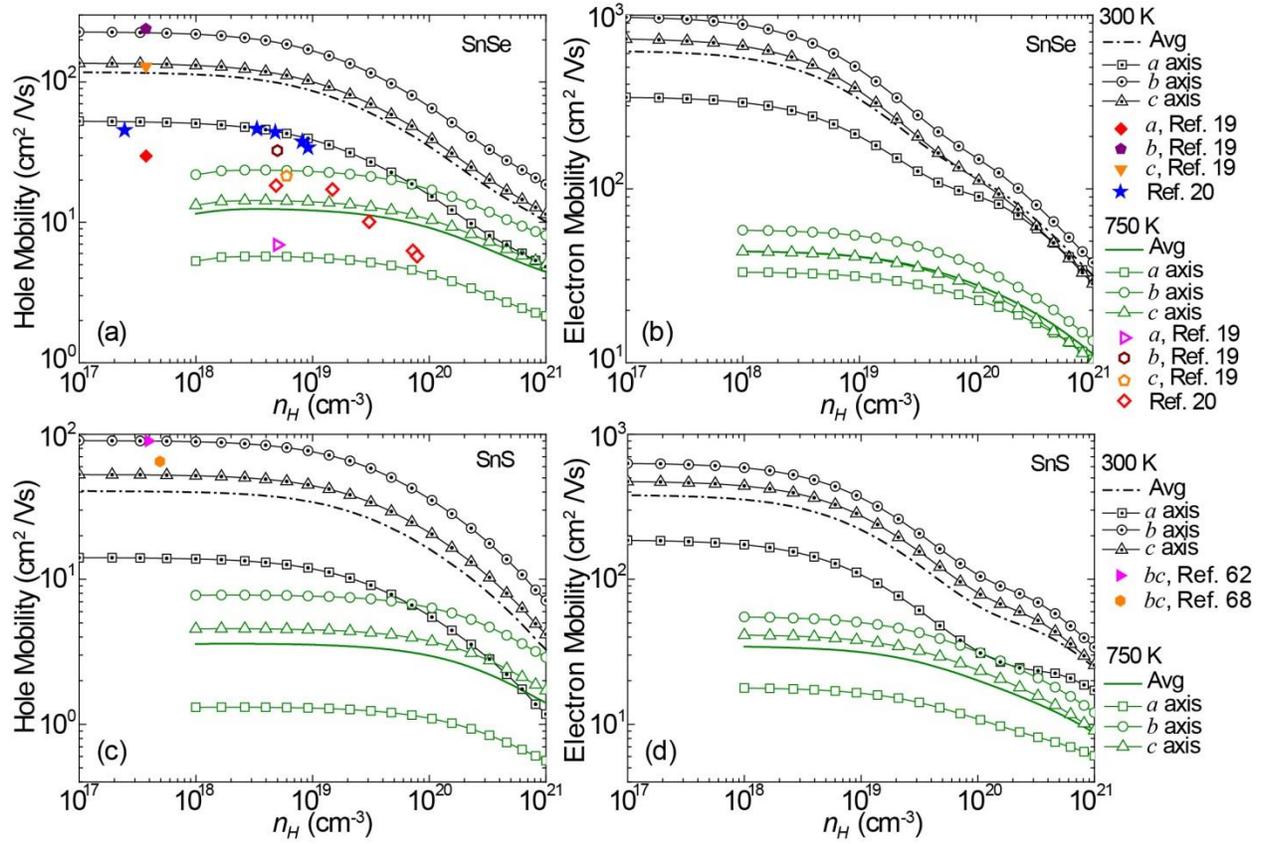

Figure 9. Hole and Electron mobilities calculated for SnSe [(a) and (b)] and SnS [(c) and (d)] as a function of carrier concentration along the *a*, *b*, *c* axes and the average ones, in comparison with the experimental results for single-[19] and poly-crystalline[20,62,68] samples.

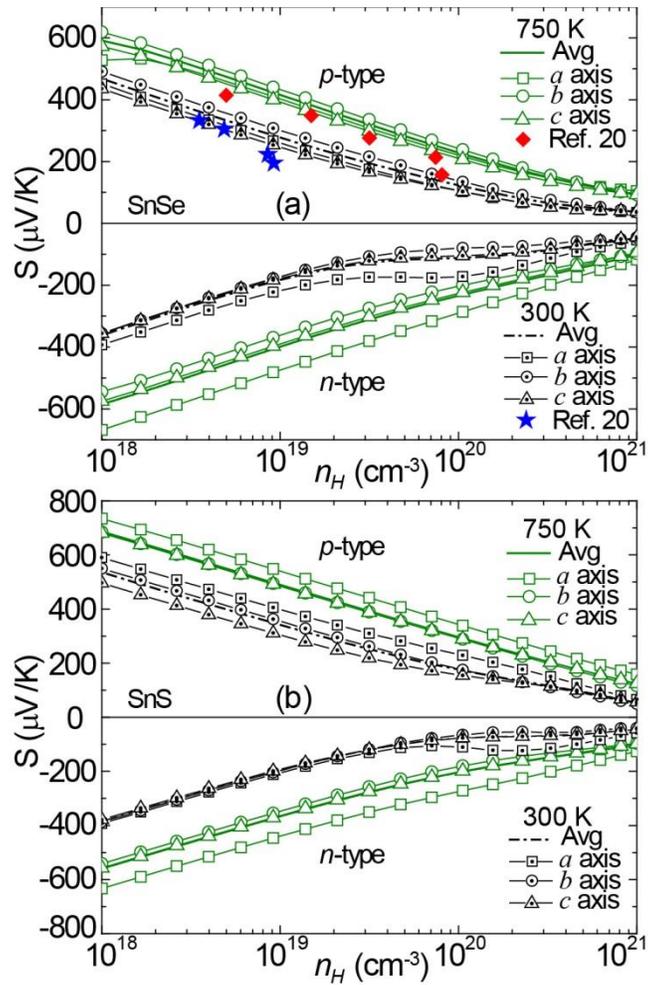

Figure 10. Calculated Seebeck coefficients as a function of carrier concentration along the *a*, *b*, *c* axes and the average ones at 300 and 750 K for *p*- and *n*-type SnSe (a) and SnS (b), in comparison with experimental results for polycrystalline *p*-type SnSe[20].

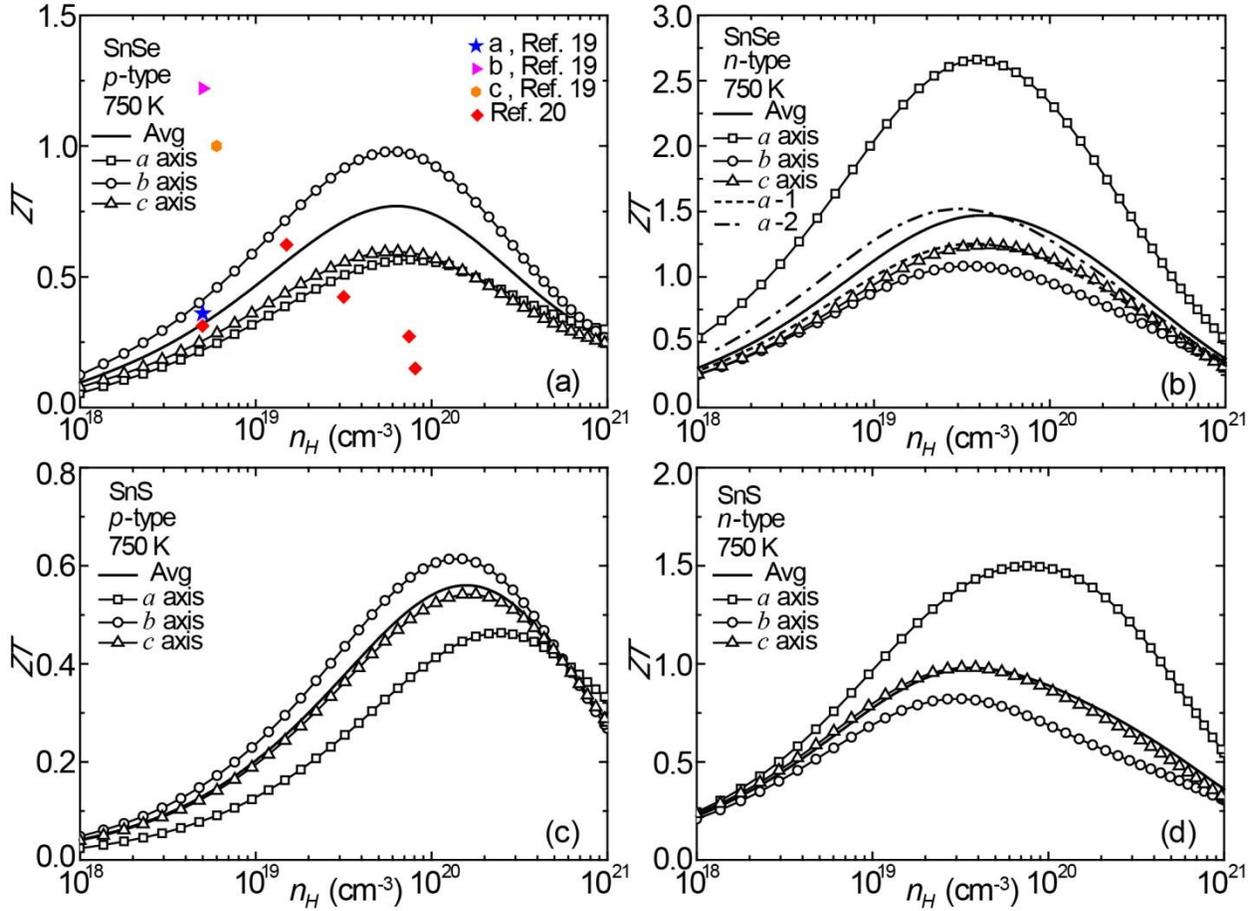

Figure 11. Calculated dimensionless figure of merits along the *a*, *b*, *c* axes and the average ones at 750 K for *p*- and *n*-type SnSe [(a) and (b)] and SnS [(c) and (d)], in comparison with the experimental results for single-[19] and poly-crystalline samples[20]. In (b), the curve "*a*-1" represents *ZT*s without considering the enhancements in $S_a$ and $\sigma_a$ while "*a*-2" is obtained by only considering the enhancement in $\sigma_a$.